\documentclass[fleqn,usenatbib]{mnras}

\usepackage[T1]{fontenc}
\usepackage{ae,aecompl}

\usepackage{graphicx}	
\usepackage{amsmath}	
\usepackage{amssymb}	

\title[CoRoT RR Lyrae stars]{
 Finest light curve details, physical parameters, and
period fluctuations of CoRoT\thanks{
The CoRoT space mission,
launched on 2006 December 27, was developed and is operated by the CNES,
with participation of the Science Programs of ESA, ESA's RSSD, 
Austria, Belgium, Brazil, Germany and Spain.} RR Lyrae stars}

\author[J. M. Benk\H{o} et al.]{
J. M. Benk\H{o},$^{1}$\thanks{E-mail: benko@konkoly.hu}
R. Szab\'o,$^{1}$
A. Derekas$^{2,1}$
and \'A. S\'odor$^{1}$
\\
$^{1}$Konkoly Observatory, MTA CSFK, Konkoly Thege M. u. 15-17., H-1121 Budapest, Hungary\\
$^{2}$ELTE Gothard Astrophysical Observatory, Szent Imre herceg \'ut 112., H-9704 Szombathely, Hungary
}

\date{Accepted 2016 August 22. Received 2016 August 19; in original form 2016 August 1}

\pubyear{2016}

\begin{document}
\label{firstpage}
\pagerange{\pageref{firstpage}--\pageref{lastpage}}
\maketitle

\begin{abstract}
The CoRoT satellite supplied the scientific community with a huge data base of variable
stars. Among them the RR Lyrae stars have intensively been discussed in 
numerous papers in the last few years, but the latest runs have not 
been checked to find RR Lyrae stars up to now.
Our main goal was to fill this gap and complete the CoRoT RR Lyrae sample.
We found nine unstudied RR Lyrae stars. Seven of them are new discoveries. We identified 
three new Blazhko stars. The Blazhko effect shows non-strictly repetitive
nature for all stars. The frequency spectrum of the Blazhko star CoRoT~104948132 contains
second overtone frequency with the highest known period ratio. The harmonic amplitude and phase
declines with the harmonic order were studied for non-Blazhko stars. 
We found a period dependent but similar shape amplitude decline for all stars. 
We discovered significant random period fluctuation for one of the two oversampled
target, CM~Ori. After a successful transformation of the CoRoT band parameters to 
the Johnson {\it V} values we  estimated the basic physical properties 
such as mass, luminosity, metallicity.
The sample can be divided into two subgroups with respect to the metallicity but otherwise the physical
parameters are in the canonical range of RR Lyrae stars.
\end{abstract}

\begin{keywords}
Stars: variables: RR Lyrae --
                stars: oscillations --
                stars: interiors --
                techniques: photometric --
                space vehicles
\end{keywords}



\section{Introduction}

The recent photometric space missions such as CoRoT \citep{Baglin06} 
and {\it Kepler} \citep{Borucki10} yielded long and quasi-uninterrupted time 
series with high precision for a huge amount of variable stars.
Although few people expected previously, many amazing
discoveries have been made for classic high amplitude variables 
stars, as RR Lyraes. Among others these mission revealed the excitation of 
low amplitude modes for many Blazhko RRab stars, one of which is responsible 
for the so-called period doubling phenomenon 
\citep{Kolenberg10, Szabo10, Kollath11}, while others show the famous 
$\sim$0.61 period ratio of the RRc stars \citep{Moskalik15} or
just appear as an additional mode(s) resulting in non-traditional double and triple
mode stars \citep{Benko10}.    

The CoRoT satellite was one of the most successful in these discoveries.
Up to now about ten peer-reviewed scientific articles published CoRoT RR Lyrae results.
The paper of \citet{Szabo14} gives a good overview of these.
Previous works discussed 13 stars observed in long runs only and till the end of the fourth 
long run towards the Galactic center direction (LRc04). Neither the short nor the late runs were 
investigated for searching RR Lyrae stars. We make up for this incompleteness with the present
work.

\section{Sample selection}\label{sec:sample}

\begin{table*}
\begin{centering}
\caption[]{Some parameters of the discussed CoRoT archive RR Lyrae stars.
The columns contain the CoRoT ID, variable name (if exists), the brightness
(specified by the Exo-Dat Catalogue \citealt{Deleuil09}),
position (RA, DEC), pulsation period $P_0$,
Fourier amplitude of the main pulsation frequency $A(f_0)$,
Blazhko period $P_{\mathrm{B}}$, observed time span $\Delta t$ in days,
and CoRoT run identifier, respectively. Last three digits of the CoRoT numbers are boldfaced.
We refer to the stars by these short numbers instead of the complete IDs.}\label{tab1}
\begin{tabular}{@{}cccccccccc@{}}
\hline
\noalign{\smallskip}
Corot~ID & Name & $m_{\mathrm{EXO}}$ & RA &  DEC&  
$P_0$ &  $A(f_0)$ &  $P_{\mathrm{B}}$      &  $\Delta t$ &  Run~ID \\
 &  & [mag] &  hh:mm:ss.ss   &  d:mm:ss.s  &  [d] &   [mag]   &   [d] &     [d]     & \\
\noalign{\smallskip}
\hline
\noalign{\smallskip}
102326{\bf 020} &     & 14.654 & 06:10:59.87 &    $\phantom{-}$4:40:32.5 &     0.77029 &      0.1296 & 79.5 &   148.31 & LRa03 \\ 
104948{\bf 132} &     & 15.217 & 18:37:43.63 &    $\phantom{-}$4:01:44.2  &     0.58642 &     0.2420  & 28.1 &   87.26  & LRc05 \\
205924{\bf 190} &     & 13.600 & 18:57:17.45 &    $-$4:16:39.4 &     0.72049 &     0.1053  &     &   20.86   & SRc02 \\
605307{\bf 902} &     & 15.170 & 06:08:42.38 &    $\phantom{-}$6:22:15.1  &     0.62418 &     0.1559 &        &     77.6 &    LRa04 \\
617282{\bf 043} &  CM~Ori   & 12.889 & 06:03:54.87 &    $\phantom{-}$8:14:32.4 &     0.65593  &    0.2859  &      &  90.48    & LRa05 \\
651349{\bf 561} &     & 14.708 & 19:14:03.69 &    $-$2:27:54.8 &     0.61179  &     0.1013  & 21.9 &   83.51  & LRc09 \\
655183{\bf 353} &     & 14.911 & 19:14:41.21 &    $-$2:06:09.1 &     0.69426  &     0.1921  &     &    83.51  & LRc09 \\
657944{\bf 259} &     & 14.954 & 19:18:08.54 &    $-$2:34:43.9 &     0.57787   &     0.1876  &     &   83.51  & LRc09 \\
659723{\bf 739} & V2042~Oph & 15.186 & 18:32:46.97 &    $\phantom{-}$7:58:05.7  &     0.53849   &     0.2761  &     &   81.22+83.47 &  LRc07+LRc10 \\
\noalign{\smallskip}
\hline
\end{tabular}
\end{centering}
\end{table*}

Since hundreds of papers appeared on the basis of the CoRoT satellite data the basic facts 
about the satellite and its equipment are well-known, we only refer here
the most important instrumental paper \citep{Auvergne09}
and on-line 
documentations\footnote{\url{
http://idoc-corot.ias.u-psud.fr/sitools/client-user/COROT_N2_PUBLIC_DATA/project-index.html}}
where the reader can find the details of the mission.
This work used the CoRoT Public Archive N2 data of Exo-fields. 
For simplicity, we used the entire data base in our   
RR Lyrae search including the initial run (IRa01), long runs (LRa01-LRa07, LRc01-LRc10), 
and short runs (SRa01-SRa05, SRc01-SRc03). Here the letters `c' and `a' denote that 
the CoRoT observed either toward the Galactic centre or anti-centre directions, respectively.
The advantage of the use of the total observing material is that we can
check our search efficiency when we compare our candidates in the early runs
with the RR Lyraes found by former studies.

\begin{figure}
\includegraphics[scale=0.31]{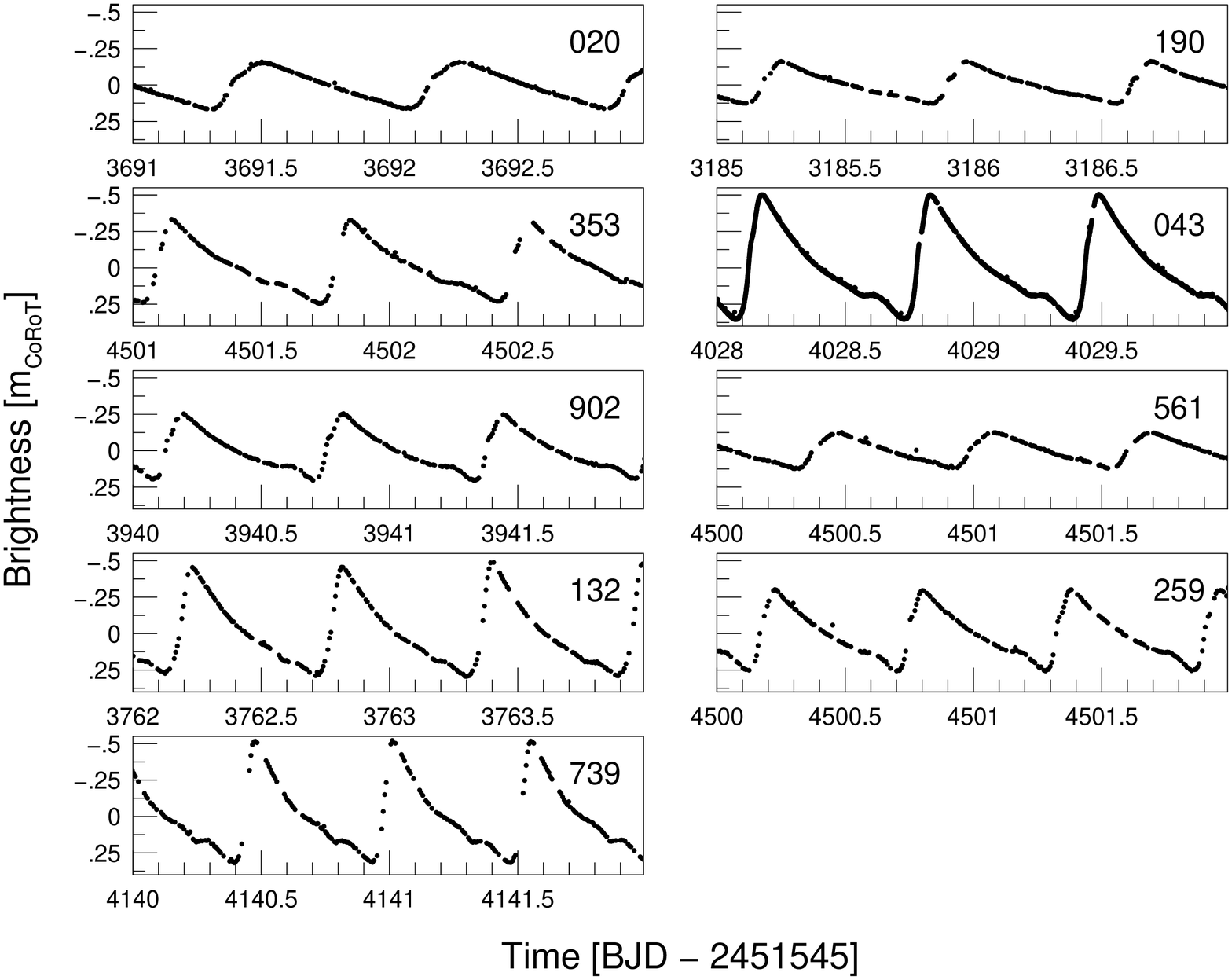}
\caption{
Light curve zoo of the CoRoT RR Lyrae stars found by this work. The time
and brightness scales are uniform for all panels. The panels
are ordered from top left to bottom right by the decreasing pulsation period.}
\label{zoo}
\end{figure}

We used the automatic CoRoT Variable Classifier (CVC; \citealt{Debosscher09}) when
we searched for RR Lyrae stars separately for each subtype in the complete data base. 
The CVC has improved a lot since the beginning.
Our present search results illustrate this well: if we decrease the probability 
level assigned by CVC of RRab stars from $>$50\% to $>$10\%, then to $>$0\%,
we found 15, 16 and 18 candidates, respectively. Investigating the candidates
by visual inspection only one of them proved to be non-RR Lyrae stars in 
the lowest probability group. All others (17) show clear RR Lyr light curve properties.
Eight RRab stars were discussed by former studies (\citealt{Szabo14}, and references therein).
Two additional RRab stars were known in the CoRoT sample (CoRoT~101315488 and CoRoT~100881648)
which have not been found by CVC. These stars are, however, heavily blended by nearby stars 
and therefore their variability have much smaller amplitude and distorted light curves 
compared to a normal RR Lyrae star. 
The CoRoT data of nine RRab stars have not been discussed anywhere yet.
These stars constitute the present sample (see Table~\ref{tab1}). 
Fig.~\ref{zoo} shows small light curve parts of each star.
 The RR Lyrae nature of two stars: CM\,Ori (CoRoT~617282043) and 
V2042\,Oph (CoRoT~659723739) have already been known but the other 
seven ones are completely new discoveries.

The same process (CVC with decreasing probability levels and
visual light curve examination) has been carried out for RRc subtype as well.
For RRc stars we get 25, 34 and 137 candidates for $>$50\%, $>$10\%, and $>$0\%
probability levels, respectively. The two known RRc stars (CoRoT\,105036241 
and CoRoT\,105735652) were already contained 
in the highest probability group. The increasing candidate numbers of the lower 
probability level groups did not result in new RRc stars. The groups
consist mostly of binary stars and some spotted and HADS/SX~Phe candidates.

\begin{figure*}
\includegraphics[angle=0,scale=.31]{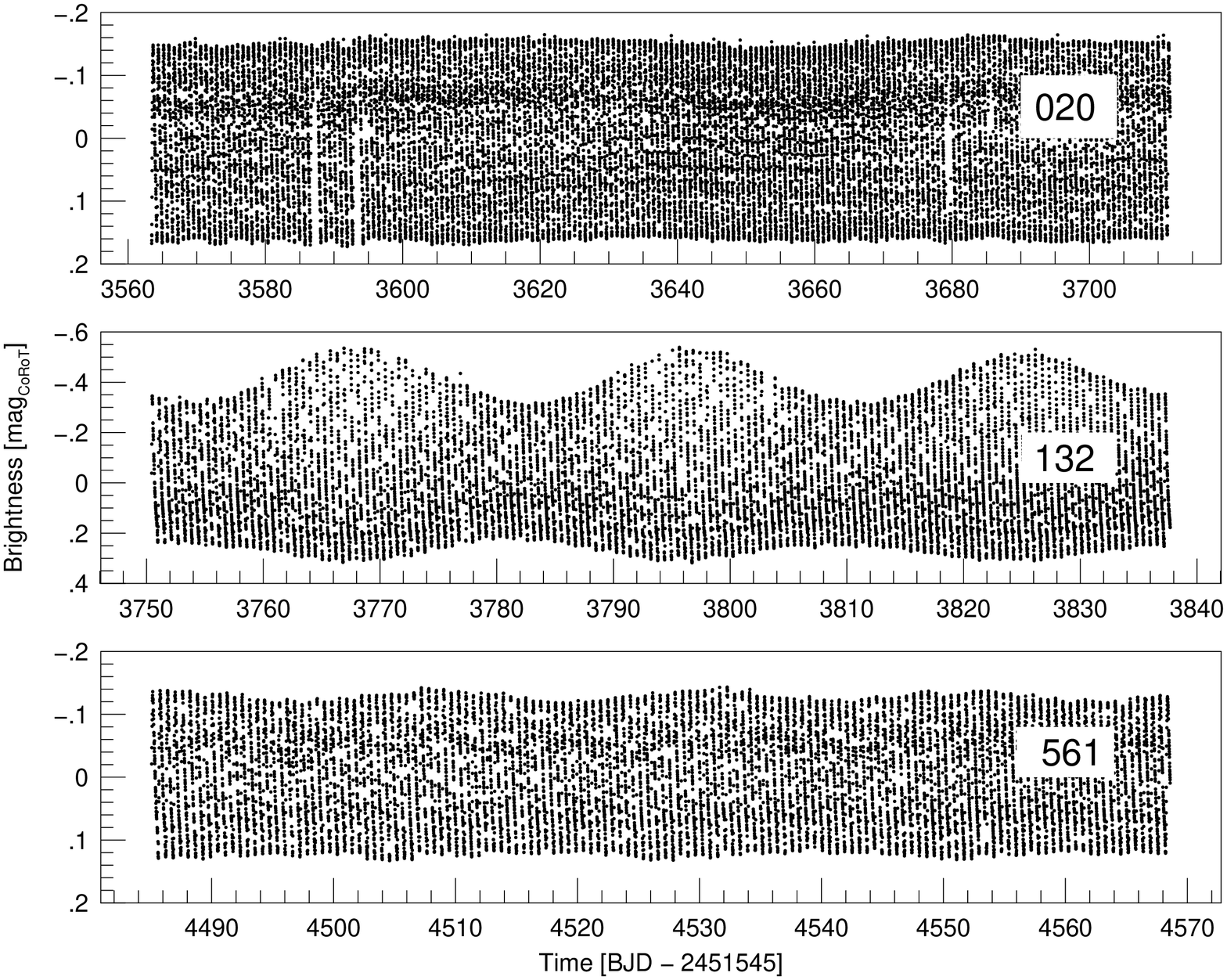}
\includegraphics[angle=0,scale=.31]{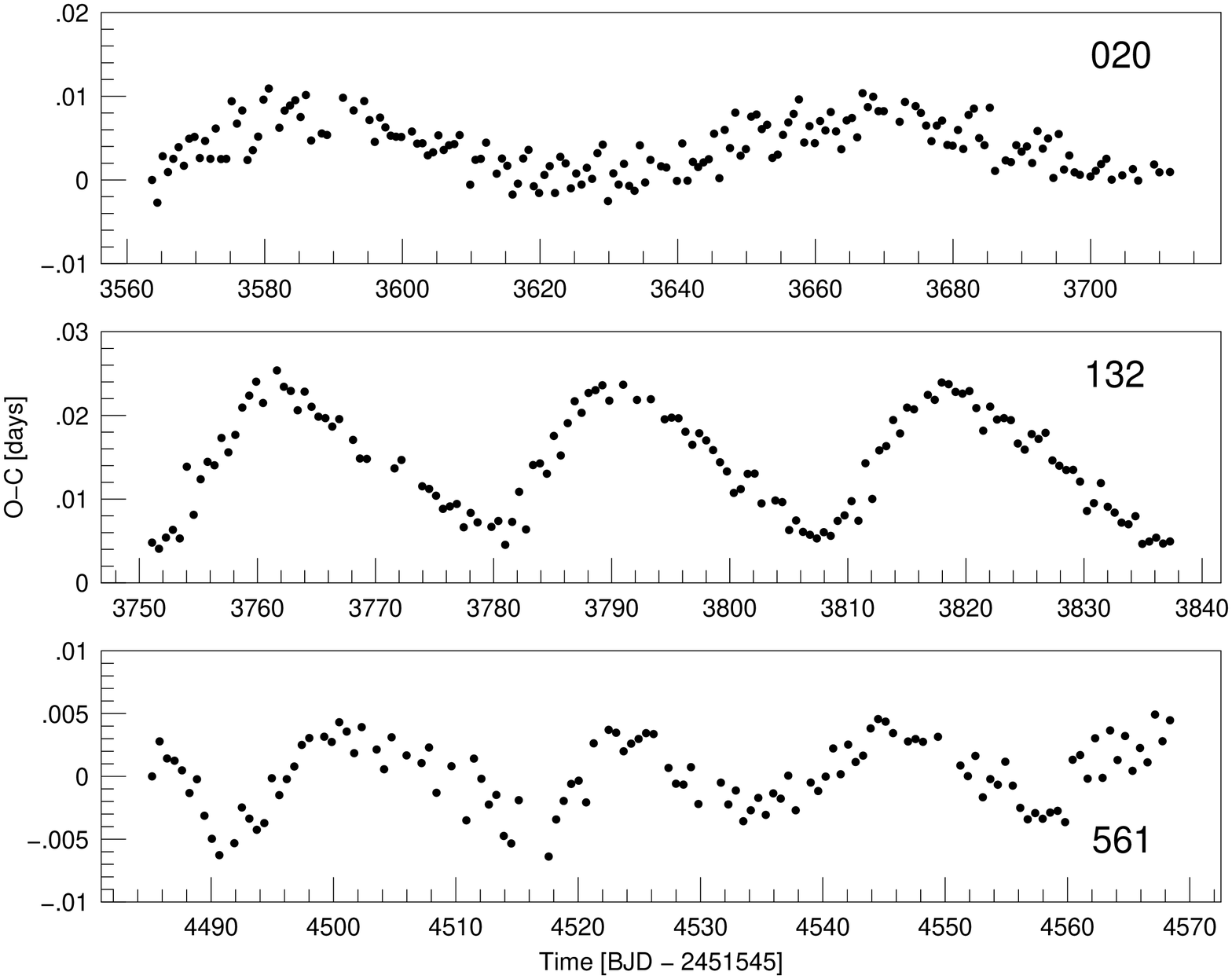}
\caption{
Complete CoRoT observations
of the new Blazhko stars (on the left) and
their O-C diagrams (on the right).}\label{Bl_fig}
\end{figure*}

In the case of double mode RR Lyrae stars (RRd) the CVC yields different
kind of results. For $>$50\% and $>$10\% probability levels we get only 
one candidate (CoRoT\,101368812) which is known as a bona fide RRd star \citep{Chadid12},
while for  $>$0\% probability level the number of candidates abruptly increases 
to 7650. We applied a semi-automatic method to handle this large number of light curves.  
We calculated the Fourier amplitude spectra in the 1\,--\,10\,d$^{-1}$
range of all the 7650 light curves.
We selected all peaks from these spectra that exceeded the 4\,$\sigma$
amplitude, where $\sigma$ was calculated as the 3\,d$^{-1}$ moving
average of the amplitude spectrum. The classical double mode RR Lyrae stars have
a well-known period ratio between their radial fundamental and first overtone modes 
$P_1/P_0\sim$0.72-0.755~d$^{-1}$ which defines the Petersen 
diagram (see e.g. in \citealt{Soszynski14}). Therefore, we selected those
objects that have their highest peak $f_\mathrm{max}$ in the 1\,--\,5\,d$^{-1}$ range and
that have at least one further significant peak in either the
$0.72f_\mathrm{max}$\,--\,$0.755f_\mathrm{max}$ or the
$f_\mathrm{max}/0.755$\,--\,$f_\mathrm{max}/0.72$ range. These are
those stars that have frequency ratio in the 0.72\,--\,0.755 range
with the largest-amplitude peak. This selection yielded 113 objects.
We investigated the light curves and Fourier spectra of these 113
objects visually. Apart from the known CoRoT\,101368812, we did not
find any further RRd pulsator. Note that 61 targets were identified as
$\gamma$\,Dor candidates and 42 as eclipsing/heartbeat binary candidates.

From now on, this paper 
refers CoRoT stars with the last three digits (boldfaced in 
the first column in Table~\ref{tab1}) of their ID number for short notation.

\section{Time series analysis}\label{sec:data}

After we extracted the downloaded data files 
we applied trend and jump filtering, outlier removal, and 
a transformation into the magnitude scale  
as we described in detail in \citet{Chadid10}\footnote{The processed 
data files are available from {\url{http://www.konkoly.hu/KIK/data_en.html}}}. 
Two of our targets (CM~Ori and \#190) were 
bright enough to be observed by CoRoT in colour mode,
but because of the uniform handling, 
we also used their integrated (white light) fluxes only.
The nominal sampling is 512 sec for all the stars, 
except CM~Ori, where the the light curve was observed with
the much denser oversampling mode (32 sec). The data set of CM~Ori
is twice as dense as the {\it Kepler}/K2 short cadence observations.

The main tool was a standard discrete Fourier
analyzer implemented by the program package MuFrAn \citep{Kollath90}.
It yields sinusoidal Fourier decomposition, therefore, we used this one
through this paper. The Nyquist frequency of the 
normal and oversampled data are 84.375~d$^{-1}$, and 1350~d$^{-1}$, respectively. 
The stability of the periods were checked by O$-$C diagrams \citep{Sterken05}.
The O$-$C diagrams were constructed from the maximum brightness times.
The proper maximum times were determined by light curve maxima with 
7-10 order polynomial fits.
 
The Fourier spectra of the analyzed nine stars are dominated 
by the main pulsation frequencies and their harmonics.
Simultaneous non-linear fits to the light curves 
with these frequencies resulted in the pulsation periods $P_0$
and the amplitudes $A(f_0)$ given in Table~\ref{tab1}. 

\subsection{Blazhko stars}

\begin{figure}
\includegraphics[angle=0,scale=.3]{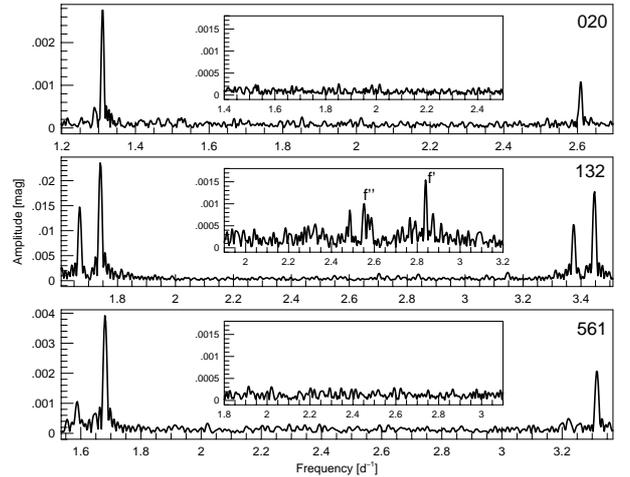}
\caption{
Fourier spectra of the Blazhko stars on the interval $(f_0-0.1, 2f_0+0.1)$,
after we pre-whitened the data with $f_0$ and all significant harmonic frequencies.
The inserts show narrower ranges $(f_0+0.1, 2f_0-0.1)$ of the spectra after a further
pre-whitening step in which we removed all the significant modulation side
frequencies as well.}\label{Fourier}
\end{figure}
{\bf \#020}: The light curve shows a slight amplitude change
(see the top left panel in Fig.~\ref{Bl_fig}). 
After we pre-whitened the data with the main pulsation frequency
and its significant harmonics we found significant peaks in the residual 
spectrum around the pre-whitened frequency positions.
We can identify them as Blazhko modulation frequencies 
as $if_0\pm f_{\mathrm B}$, ($i$ is positive integer). 
The Blazhko period $P_{\mathrm B}$ given in Table~\ref{tab1} was calculated from the average 
of the side peak differences around the main peak. 
The side peak amplitudes are highly asymmetric, the right hand side peaks are much
higher than the left hand side ones (see the top panel in Fig.~\ref{Fourier}). 
This asymmetry, however, decreases
with the increasing harmonic order: while the $f_0-f_{\mathrm B}$ and
$f_0-2f_{\mathrm B}$ are around the detection limit, $f_0-3f_{\mathrm B}$
and even $f_0-4f_{\mathrm B}$ are much less asymmetric. 
This behaviour has a natural explanation in the simultaneous 
amplitude and frequency modulation framework \citep{Benko11}. 
In the O$-$C diagram (top right in Fig.~\ref{Bl_fig}) we indeed find
a periodic signal with 80.7~d$^{-1}$ period and $\sim$5~min 
amplitude which can be identified as the frequency modulation 
part of the Blazhko effect.

The pulsation period of the star \#020 is one of the longest ($P_0=0.77029$~d) 
ever found for a Blazhko RR Lyrae star. 
Its Blazhko period is long enough to test the relation between the pulsation
period and Blazhko period discovered by \citet{Jurcsik05}.
They found that the Blazhko frequency, which was represented by the
frequency separation between the main pulsation frequency and the side
frequencies, has a maximal allowed value at a given pulsation frequency. These maximal 
values form a linear upper envelope curve in the pulsation frequency vs. Blazhko frequency
diagram  (see fig.~6 in \citealt{Jurcsik05}).
 The position of \#020 is under this upper envelope line supporting the finding of the paper.
Both the amplitude and frequency modulation parts of 
the two observed Blazhko cycles differ from each other (top panels in Fig.~\ref{Bl_fig}).
This raises the possibility of multiperiodic (or chaotic) nature of the modulation,
but such a short observing span does not allow us to confirm or reject these hypotheses.

{\bf \#132}: The light curve in the middle left panel in Fig.~\ref{Bl_fig}
shows three clear Blazhko cycles with decreasing amplitudes. The Fourier spectrum
contains evident triplet structures around the main pulsation frequency and all its harmonics.
The low-frequency range of the spectrum contains three significant frequencies: 
$f^{(1)}=0.0355$~d$^{-1}$, $f^{(2)}=0.0143$~d$^{-1}$ and 
$f^{(3)}=0.0475$~d$^{-1}\approx f^{(1)}+f^{(2)}$.
The frequency $f^{(1)}$ can be interpreted as the Blazhko frequency $f_{\mathrm B}$ itself,
the $f^{(2)}$ most probable belongs to the total time span, and $f^{(3)}$ is a linear
combination of these two frequencies.  
After the first two pre-whitening steps in which we removed the main pulsation frequency,
its harmonics and all significant Blazhko side peaks the residual spectrum shows two 
groups of significant peaks between
the harmonics (middle panel in Fig.~\ref{Fourier}). 
These peaks are most robust between the main pulsation frequency and the first harmonic.
The highest peak is at $f'=2.8394$~d$^{-1}$ while the second one is at $f''=2.5519$~d$^{-1}$.
The latter frequency can be identified as $f''\approx 3/2f_0$ which belongs to the period 
doubling phenomenon. What about the frequency $f'$? Numerous studies 
\citep{Benko10, Benko14, Poretti10, Szabo14, Molnar15}
 found frequencies at this position: between the  period doubling (PD) frequencies and
the first overtone. These frequencies are explained by the $f_2$ second radial overtone mode
or a non-radial mode which is excited close to the position of the second overtone mode. 
\#132 has the highest period ratio $P_2/P_0=0.601$ ever detected.

\begin{figure}
\includegraphics[angle=270,scale=.33]{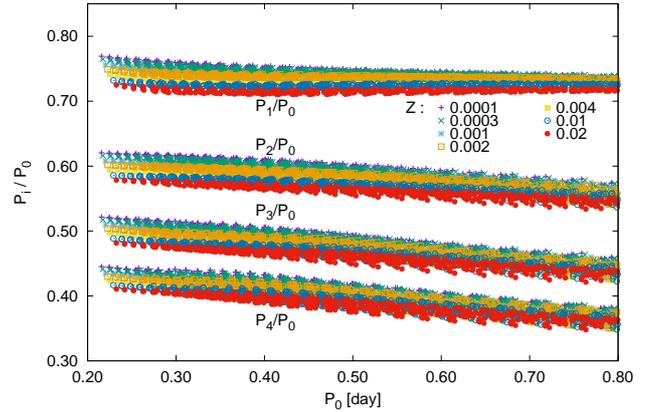}
\caption{
Theoretical Petersen diagram based on linear convective RR Lyrae models
up to the 4th radial overtone modes. The symbols denote different metallicities.}
\label{dm_model}
\end{figure}
The identification of the $f'=f_2$ is verified with theoretical model calculations.
Fig.~\ref{dm_model} shows a theoretical Petersen diagram where the periods 
$P_0$, $P_1$, $P_2$, $P_3$, $P_4$ belong to
the radial fundamental, first, second, third and fourth overtone modes, respectively.
The different metallicity models are noted by different symbols. The parameters and
details of the models are the same as it was described in \citet{Chadid10}.
As we see the ratio of $f'$ is in the allowed range of second overtone frequencies. 
That is the Fourier spectrum of  \#132 shows the PD  and the radial second overtone mode
frequencies simultaneously similarly to some other stars in the CoRoT and {\it Kepler} samples 
(e.g. CoRoT~ 101128793 \citealt{Poretti10}; V355~Lyr, KIC~7257008, and KIC~9973633 \citealt{Benko14}). 

The O$-$C diagram in the middle right panel in Fig.~\ref{Bl_fig} shows evident
frequency modulation. The Fourier spectrum of the O$-$C diagram contains $f_{\mathrm B}$
and $2f_{\mathrm B}$ reflecting the non-sinusoidal nature of the variation.
The consecutive Blazhko cycles are different but do not show declining amplitudes 
as opposed to the light curve. The strictly mono-periodic nature of the
Blazhko effect, similarly to the case of the star \#020, 
 can be ruled out both on the basis of the light and O$-$C curves.
 
{\bf \#561}: Both the light and O$-$C curves (bottom panels in Fig.~\ref{Bl_fig}) 
show the Blazhko effect. The amplitude modulation seems to be more regular than the
frequency modulation part. The modulation side peaks in the Fourier spectrum
of the light curve are asymmetric: left hand side peaks have lower 
amplitudes than the right hand side ones. After we removed all the 
significant side frequencies, no additional frequencies have been found in the residual spectrum.
Interestingly, while all stars that show additional frequencies 
show the Blazhko effect as well \citep{Benko15}, 
the reverse statement is not true. There are known some 
CoRoT \citep{Szabo14} and {\it Kepler}/K2 \citep{Benko14, Molnar15} 
Blazhko stars without any additional frequencies. As we show here the stars \#561
and \#020 seem to belong to this group.
We can speculate whether this finding is a selection 
effect or a real feature. The selection effect explanation is
supported by the time dependence of these  additional frequencies. 
They have low and highly time dependent amplitude. So these frequencies 
could disappear from the frequency content for a while and then emerge again.

\subsection{Non-Blazhko stars}

\subsubsection{Stability and change of the pulsation period} 

\begin{figure}
\includegraphics[scale=0.3]{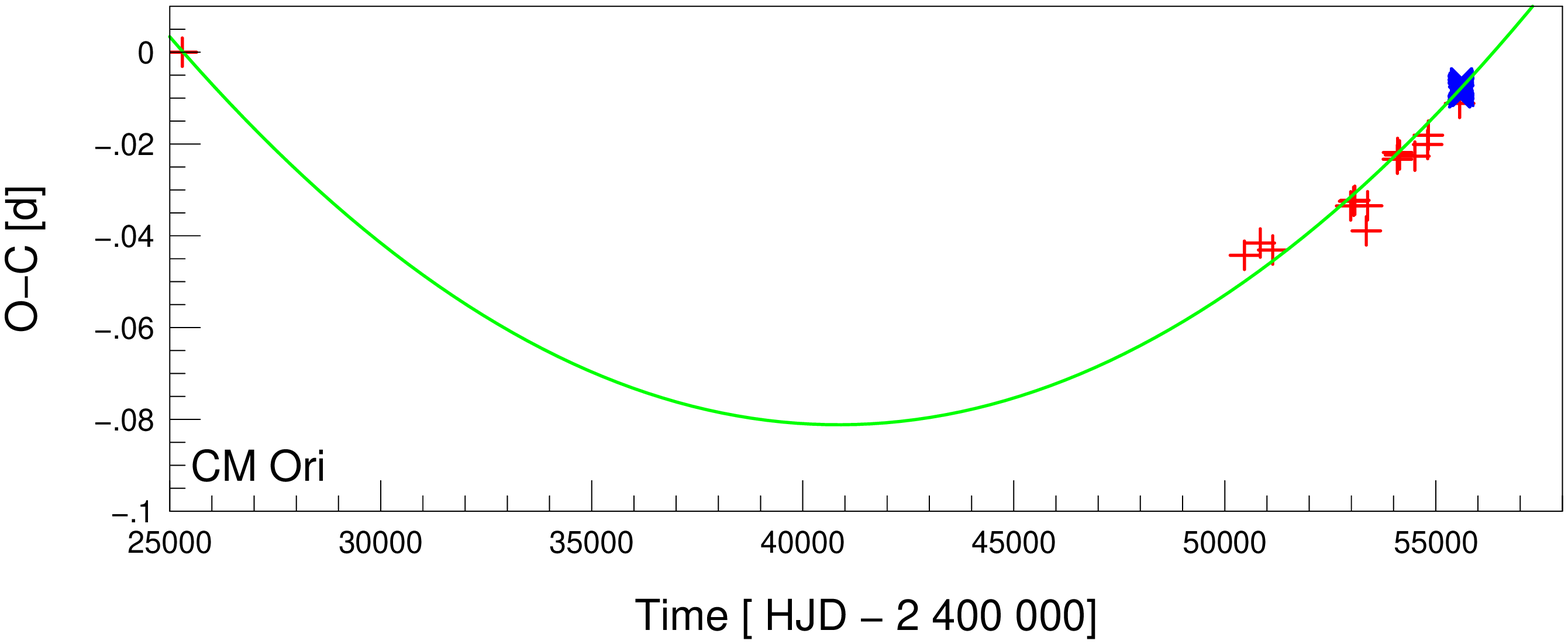}
\includegraphics[scale=0.3]{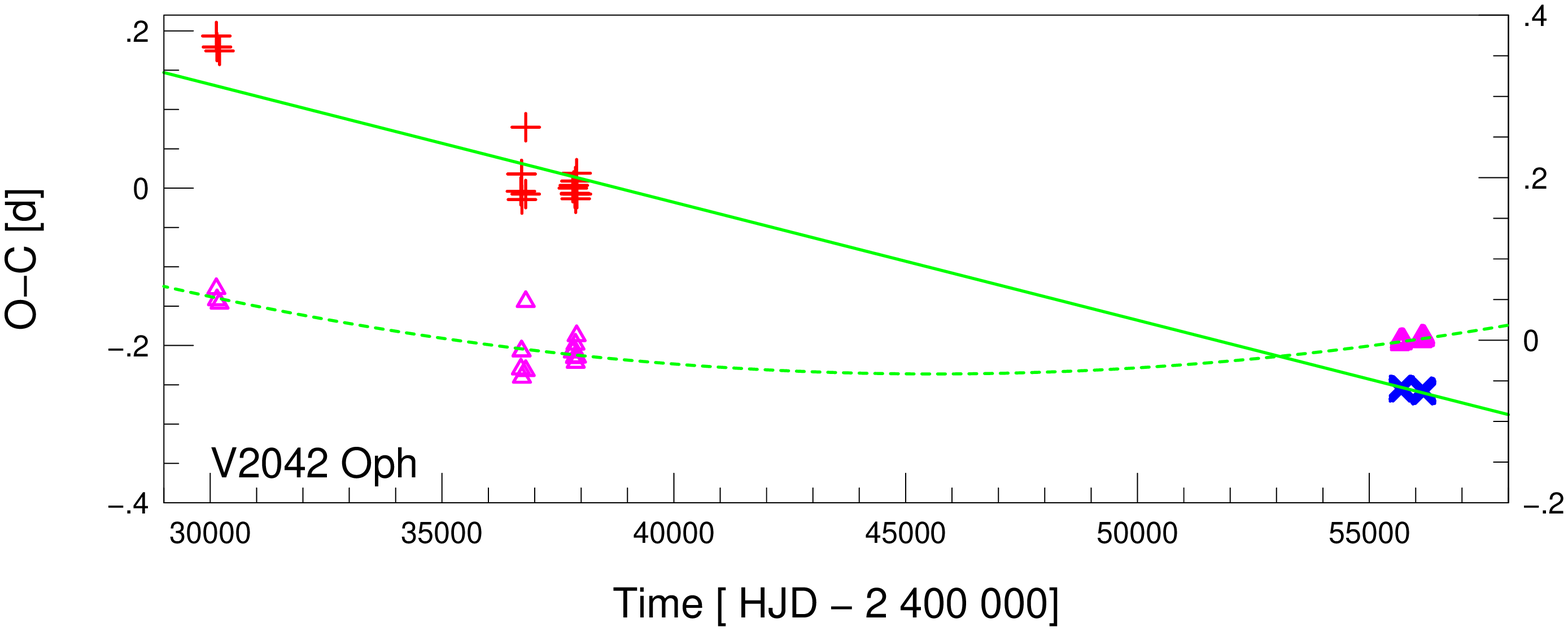}
\caption{
Long time-scale O$-$C diagrams of CM~Ori (top) and V2042~Oph (bottom).
The (red) plus symbols denote the historical data the (blue) x
signs are the CoRoT measurements. The green continuous
curve in the top panel shows the quadratic fit to the data.
The (purple) triangles in the bottom figure mean the residual O$-$C data 
after we removed the linear trend showed by green continuous line. 
The right hand side vertical scale in the bottom panel belongs to the
residual values, while the dashed curve shows the quadratic fit 
to theses data
\label{O-C_long}} 
\end{figure}
As we noticed two of our stars (CM~Ori and V2042~Oph) 
were known RR Lyrae variables. CM~Ori was discovered early
by \citet{Ross25} while V2042~Oph was found in the Sonneberg
plates by \citet{Hoffmeister49}. The stars were  
targets of maxima observation from time to time. The GEOS RR Lyrae data 
base\footnote{\url{http://rr-lyr.irap.omp.eu/}} \citep{LeBorgne07} contains  
13 and 17 maxima values for CM~Ori and V2042~Oph, respectively.
We completed the data set of CM~Ori with three new maxima observed by TAROT
Survey \citep{LeBorgne12} and the current CoRoT observations, while 
we did not use maxima of \citet{Hoffmeister30}
because of their large uncertainties ($\pm$7~min).
The constructed long-term O$-$C diagrams can be seen in the top panel in Fig.~\ref{O-C_long}.
The O$-$C points distribute along a parabolic function suggesting a
continuous period increase. We fitted the simple quadratic formula
\[
\mathrm{O}-\mathrm{C}=a t^2 + b t + c, 
\]
where $a, b, c$ are constants, $a=\beta$ is the linear period change rate, $t$ is the 
time\footnote{We mention here that the GEOS data base provides Heliocentric Julian date while
CoRoT uses Baricentric Julian date. The difference of these two types of dates are
very tiny and therefore we neglected it.}. The rate of the period increase is
$\beta=3.37\cdot 10^{-10} \pm 6 \cdot 10^{-12}$~dd$^{-1}$, or, $0.092 \pm 0.002$~dMy$^{-1}$.

The O$-$C diagram of V2042~Oph (bottom panel in Fig.~\ref{O-C_long}) consists of
the GEOS data and points from the two CoRoT runs. We shifted the 512~sec CoRoT date
with 224~sec as it was described by \citet{Weingrill15}. The  O$-$C diagram 
suggests that the GEOS period ($P_0$=0.5385~d) is too long. 
From the slope of a fitted linear 
we get $1.5\cdot 10^{-5}$~d correction, which
is good agreement with our new CoRoT period ($P_0=0.53849$). Applying this correction
(viz. subtracting the fitted linear shown by the continuos line) 
we detect a period increase again. From the quadratic fit (dashed curve) the period change rate is 
$\beta=3.9\cdot 10^{-10} \pm 2 \cdot 10^{-11}$~dd$^{-1}$, or, $0.11\pm 0.005$~dMy$^{-1}$.

These period change rates are in agreement with the canonical
stellar evolution models \citep{Dorman92, Demarque00, Girardi00} 
which predict the red-ward evolution rate 
between $\alpha=1$ and 10$\cdot 10^{-10}$~dd$^{-1}$, where $\alpha=\beta /P_0$.
In our cases $\alpha=5.12$, and 7.2$\cdot 10^{-10}$~dd$^{-1}$ for CM\,Ori and
V2042\,Oph, respectively. These actual values are also fit well to the empirical
rate distributions determined for the globular cluster M3 by \citet{Jurcsik12}.

After the investigation of the long term period changes we turn to 
the cycle-to-cycle stability of the non-Blazhko stars. \citet{Nemec11} 
performed a stability analysis for the {\it Kepler} sample. They calculated 
the time dependence of the main Fourier parameters, namely
$\varphi_1(t)$, $\varphi_{31}(t)$, $A_1(t)$, $R_{21}(t)$. All these functions 
proved to be constant with a small random scatter. \citet{Derekas12}
followed a different approach when they found random period fluctuation in
the {\it Kepler} Cepheid V1154\,Cyg. They studied the O$-$C diagram, which 
suggested the cycle lengths scatter of 0.015-0.02~d ($\approx$20-30~min), 
meaning $\approx 0.3\%$ of 
the pulsation period. Similar magnitude of cycle-to-cycle light
curve fluctuation would be 1-2 minutes for a typical RR Lyrae stars.
We have a chance to detect such a small time-scale effect only for the oversampled 
stars. Two CoRoT RRab stars CM Ori and CoRoT\,103800{\bf{818}} \citep{Szabo14}
were observed in oversampled mode. Here we investigate the
data sets of these two stars in details.

\begin{figure}
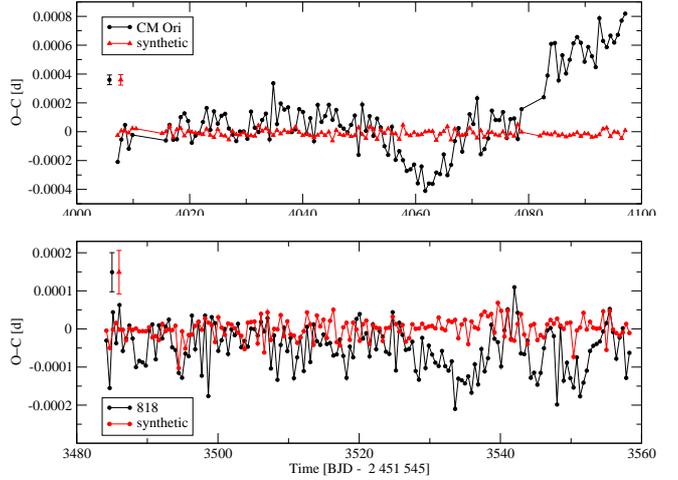

\includegraphics[angle=0,scale=.31]{043-comp}
\includegraphics[angle=0,scale=.31]{818comp}
\caption{
O$-$C diagrams of CM~Ori (top) and \#818 (bottom) constructed from the 
phase shifts. Black dots represent the observed data, red triangles denote the 
synthetic data (see the text for the details). For better visibility we 
show the error bars as separated symbols and we connected the consequtive points.
We call the attention of the reader to the different vertical scales.
\label{O-C_short}}
\end{figure}

First we constructed the traditional O$-$C diagrams from the light curve maxima.
Both diagrams are constant lines with a standard deviation of $\sigma=0.00061$~d,
and 0.00132~d for CM Ori and \#818, respectively. In the case of CM Ori this
scatter means 0.88~min, or, 0.09\% of the period, while for
\#818 these values are 1.9~min and 0.3\%. To test the accuracy of this
O$-$C method we prepared synthetic light curves using the 
Fourier solutions (amplitudes and phase) of the real stars and adding to them Gaussian noise.
The noise was set that the RMS of the synthetic light curve fits became equal to
the measured light curves: 0.0037~mag for CM Ori, and 0.0068~mag for \#818, respectively.
The standard deviation of the O$-$C diagrams of these stationary periodic synthetic data
are $\sigma=$0.00058~d (CM Ori) and  $\sigma=$0.00121~d (\#818). Since we recovered 
the observed scatters our conclusion is that the random fluctuation of the pulsation period is
lower than the detection limit of this method.

Secondly a more sensitive method was used to measure phase shifts of each pulsational 
cycle and transform this to an O$-$C diagram. This approach uses the 
complete light curve instead of small parts around the maxima and could result in
more precise O$-$C values than the traditional method.
The method is the same as it was successfully 
used by \citet{Derekas12} for detecting the random period changes of
the {\it Kepler} Cepheid V1154~Cyg. A template curve was defined 
by fitting 36th order Fourier polynomial to the phase diagram of
one pulsational cycle. Then we used it -- allowing only
vertical and horizontal shifts  -- to fit each phase in the light curve.
Combining the phase shifts and the period with the epoch, we 
calculated the O$-$C values 
which are plotted in Fig.~\ref{O-C_short} with black dots. The same process
has been done for the above mentioned synthetic data as well. The results
are shown in  Fig.~\ref{O-C_short} with red triangles. The red and black symbols 
show different curve shapes for both stars. This difference is significant
for CM~Ori (top panel in Fig~\ref{O-C_short}) but it is also less than
0.0008~d (1.2~min). What is the reason of this deviation?
A frequency modulation (Blazhko effect) is unlikely because no
side peaks have been detected around the harmonics in the Fourier spectrum. 
The rather irregular shape of these O$-$C curves rule out the binarity explanation.
The most natural explanation is the intrinsic random period fluctuation.
In other words: the RR Lyrae stars are also not precise clocks.
Due to the cumulative nature of the O$-$C diagram we see here a random walk.
A detailed discussion of the phenomenon and its appearance in O$-$C diagrams 
is given in \citet{Koen05}.   

A possible irregular period jitter in Cepheids and RR Lyrae stars was
proposed by  several authors \citep{Sweigart79, Deasy85, Cox98} 
on different theoretical grounds.
But up to now it has not been detected yet for RR Lyrae stars.
It is possible that our detected small period changes of CM~Ori are
typical. The 1.2~min cumulative period change means no more than
1-2 seconds differences between the length of the subsequent pulsation cycles.
Such a small difference could be detect only in precise uninterrapted high 
cadence data.  

\subsubsection{Amplitude and phase behaviour of the harmonics}\label{Sec:amp_phi}

\begin{figure*}
\includegraphics[angle=0,scale=.31]{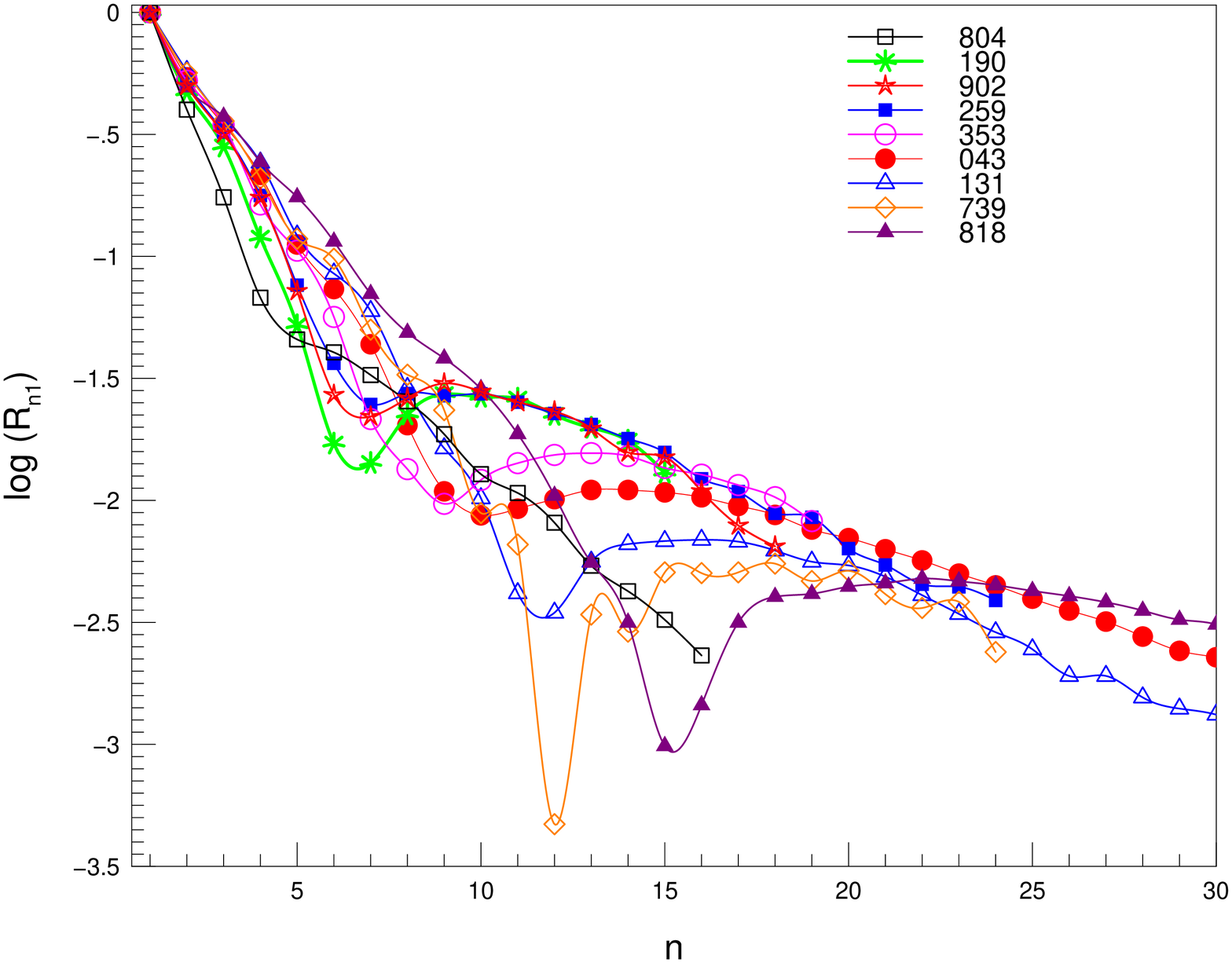}
\includegraphics[angle=0,scale=.31]{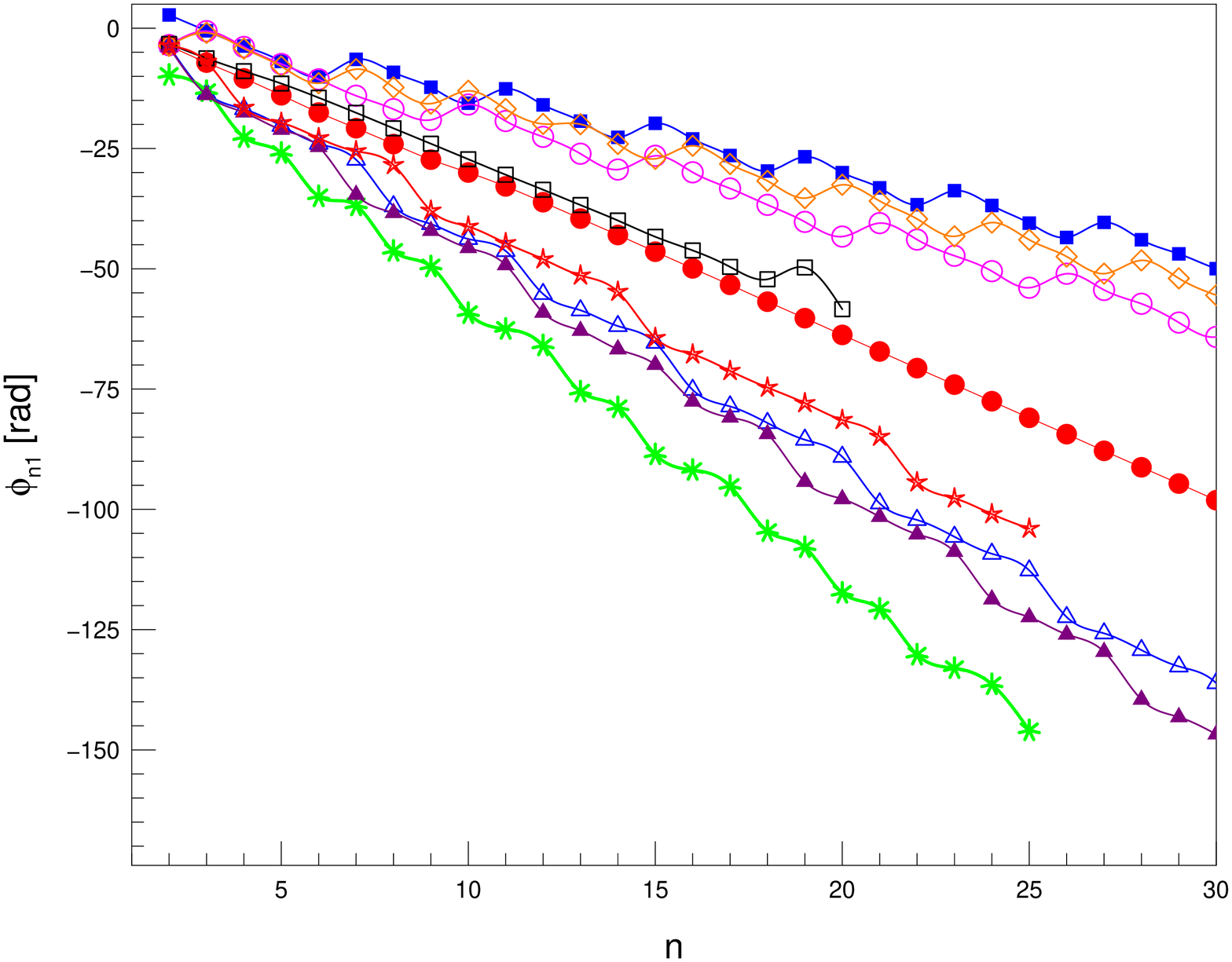}
\caption{Left panel Amplitude ratio $R_{n1}$
distribution of the non-Blazhko stars vs. harmonic order $n$.
Each symbol represents an amplitude ratio of a given star.
The amplitude ratios belonging to a different star denoted by different symbols (see 
the legend in the top right corner of the figure).
The consecutive amplitude ratios of a given star are connected with
 a continuous line for the better visibility. Right panel Epoch independent 
phase differences $\varphi_{n1}$ vs. harmonic orders $n$. As opposed to 
the common handling the phase differences
are not transformed to a restricted interval (e.g. $0<\varphi_{n1}<2\pi$),
for the sake of the better visibility.}\label{amp}
\end{figure*}
\begin{figure}
\includegraphics[angle=0,scale=.31]{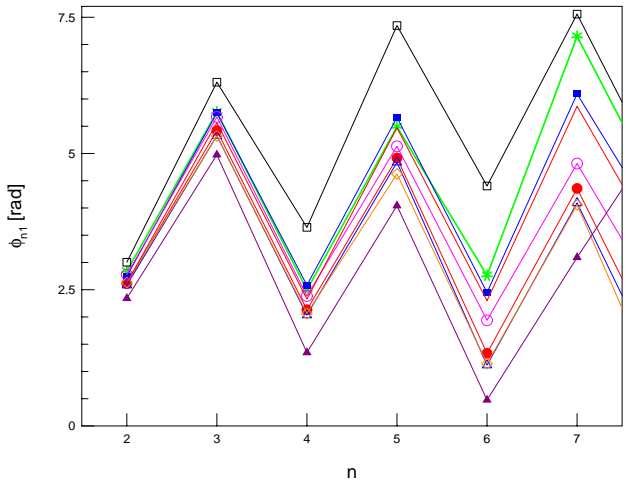}
\caption{
The low order phase differences
plotted on the usual scale. The designations are the same
as in the right hand side panel of Fig.~\ref{amp}.
}\label{phase_traditional}
\end{figure}
The Fourier amplitude of the harmonics decreases with the harmonic order.
Previously this decline was believed to be exponential but (for 
at least non-Blazhko stars) it was not studied.
Satellite data which provide us high number of significant harmonics showed
that this nearly exponentially decay can not necessarily be described by a 
single exponential function. Both of the studied non-Blazhko RRab stars 
(CoRoT\,101370131, \citealt{Paparo09}, CoRoT~103800818 \citealt{Szabo14})
show similar behaviour: their amplitudes decline exponentially for a while then this
decline shows a bump then the amplitude
decreases again exponentially but with a smaller exponent than with the lower order
harmonics. In this section we investigate how general this feature is.

We mention here that the amplitude decline of the Blazhko stars were investigated
in several papers (e.g. \citealt{Smith99, Jurcsik05b, Jurcsik06, Jurcsik09}).
These works concentrated mostly on the different declines of the harmonics and side 
peak amplitudes. Recently \citet{Zalian16} found a hyperbolic fit for the 
harmonic decline to be better than the exponential ones.
In the case of Blazhko stars, however, the amplitudes of the harmonics are heavily 
affected by the frequency/phase
modulation part of the Blazhko effect \citep{Benko11}. In other words, in the amplitude
decline behaviour the basic physics of the pulsation and unknown aspects of 
the Blazhko effect are mixed. 
To avoid this complication we restricted our study to the non-Blazhko stars.

Our sample contains six non-Blazhko RR Lyrae stars. 
We completed this sample with the three formerly discussed non-blended CoRoT
stars (CoRoT~101370{\bf{131}}, CoRoT~103800{\bf{818}}, CoRoT\,104315{\bf{804}}, \citealt{Szabo14}).
The used Fourier parameters $A_i$ and $\varphi_i$ of these stars are given in an 
electonic-only table\footnote{The table contains all stars' data consecutively.
In the three columns are the frequency in d$^{-1}$, the amplitude in CoRoT magnitude, and the phase in radian.}.

If we plot the Fourier amplitude ratios $R_{n1}=(A_n/A_1)$  as a function of the harmonics $n$ 
for these nine CoRoT non-Blazhko stars we get Fig.~\ref{amp}. In this logarithmic plot
we see that the monotonic decay holds to the 5-15th harmonic order.
The exact value varies from star to star. The distribution curves seem to build up 
a sequence from the stars \#804, where only a slight break appears at the 5th harmonic, 
to the star \#818, where a deep minimum exists at around the 15th harmonic.
It can also be realized that the depth of the dip tends to increase with the increasing
harmonic order. This trend seems to break at \#739, but its dip is defined by
only one small (marginally significant) amplitude of the 12th harmonic frequency,
the extraordinarily deep minimum might cause this uncertainty. 
The order of stars settled by the diagram in Fig.~\ref{amp} is almost the same as the order of
decreasing pulsation period. The longest period non-Blazhko star of the sample is
\#804 ($P_0=0.7218221$~d), while the shortest one is the star \#818  ($P_0=0.4659348$~d).
Seven stars from the nine one follow this period length order. The two exceptions are 
\#902 and \#259 which are located in longer period positions of the sequence than their
real period.
 
The Fourier amplitudes and phases collectively contain the complete information of a signal.
Although this is well-known, the phase spectra are studied much less frequently than the 
amplitude ones. In the right-hand-side panel of Fig.~\ref{amp} we show the 
epoch independent phase differences $\varphi_{n1}$ of the nine 
studied RRab stars. 
The panel shows the phase differences according to the definition: 
$\varphi_{n1} = \varphi_{n} - n \varphi_{1}$ ($n$ integer), where we did not transform the
differences to a restricted interval as it is common. The reason is illustrated with the 
Fig.~\ref{phase_traditional} 
where we transformed the first seven $\varphi_{n1}$ quantities into the $(0, 5/2\pi)$ interval. 
Here the $\varphi_{n1}(n)$ distributions of different stars are mixed. The distributions
are hardly traceable. 

By looking at the right panel in Fig.~\ref{amp}
we notice that these $\varphi_{n1}(n)$ distribution functions are deviate from each 
other. The distance of the nearby functions increases with the increasing 
harmonic order forming a fan shape structure on the  $\varphi_{n1}(n) - n$ surface. 
This spread suggests that the fine structure of the 
individual stars' light curve differ more than their main character described with
the low order harmonics. The shape of these  $\varphi_{n1}(n)$ decline functions 
vary form the middle of the fan-shape distribution to the wings: 
from the pure linear (\#043) to intensely crisscross shapes (e.g. \#259, \#190), 
respectively. 

The fine structures of the light curves such as the amplitude distribution 
of the higher harmonics have never been studied theoretically.
\citet{Stellingwerf87} estimated Fourier parameters $R_{n1}$ and $\varphi_{n1}$ for $n\le 10$
using a simple non-adiabatic one-zone pulsation model. They did not investigate the amplitude ratio 
and phase difference distributions with respect to the harmonic orders, but 
their figures fig.~6-7. suggest monotonic declining amplitudes with the increasing $n$,
except some model light curves, where their `acuteness' parameter is small. In these cases 
the higher order amplitudes ($n\ge 7$) could be higher than the lower ones.
The phase plots in their fig.~8-9. predict monotonic increasing phase differences (within
the interval $0\le \varphi_{n1} \le 3\pi$) at least 
until $n\le 6$, which is evidently not the case for the observed light curves
(see Fig.~\ref{phase_traditional}).  
The amplitude ratios and phase differences of theoretical light curves 
were investigated by \citet{Dorfi99} using a more enhanced non-linear convective 
1D pulsation approach but their calculations extended up to the 5th harmonic order only. 
The more realistic physical treatment gives more realistic predictions for
phase differences 
(see fig.~8 in the paper)\footnote{The phases used for 
that plot coming from a cosine decomposition. 
They need to shift (e.g. $\varphi_{21}=\Phi_{21}-\pi/2$, 
 $\varphi_{31}=\Phi_{31}+\pi$) for a direct comparison to our work.}. 
Reading the related values from the different panels 
we can recognize a similar crisscross behaviour
of the phase difference distribution as we see in our Fig~\ref{phase_traditional}.

\begin{figure}
\includegraphics[angle=0,scale=.31]{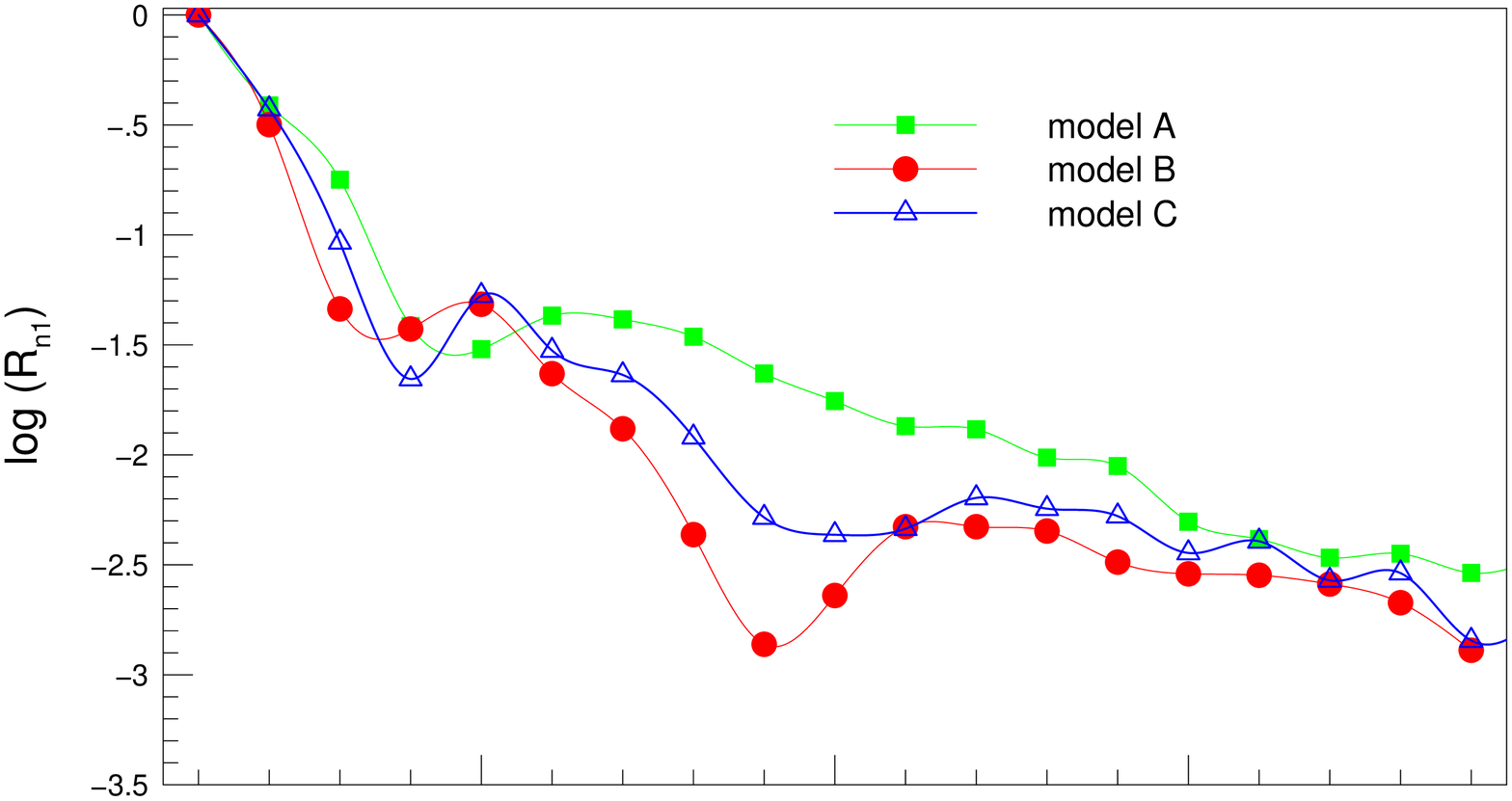}
\includegraphics[angle=0,scale=.31]{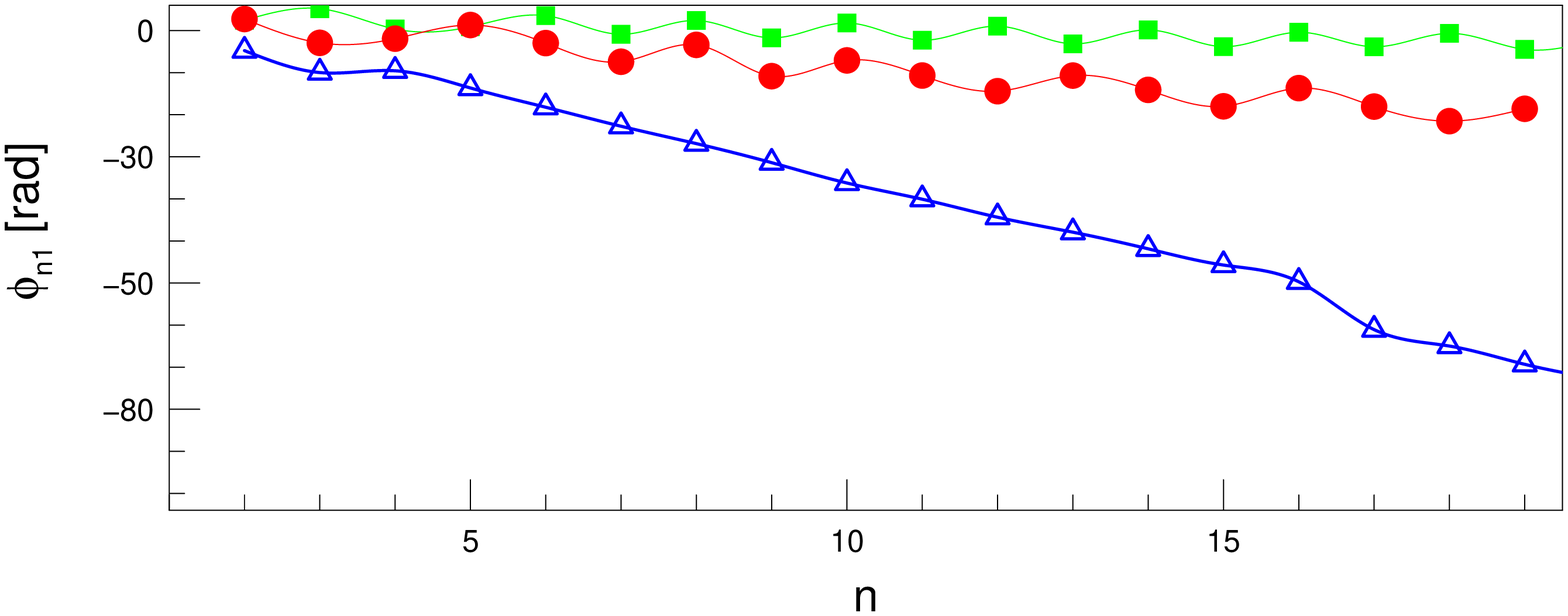}
\caption{
Amplitude ratio distribution $R_{n1}(n)$ (top panel) and epoch independent 
phase differences $\varphi_{n1}$ vs. harmonic orders $n$ (bottom panel) 
of some synthetic light curves.}\label{amp_phase_modell}
\end{figure}
We compared the Fourier amplitude and phase distribution of the 
observed and model light curves calculated from the 
photospheric temperature $T(t)$ and gravitational acceleration
$\log g(t)$ curves by the Florida-Budapest hydrocode \citep{KollathBuchler01, Kollath02}.
Since the models use the exact black-body condition we can use
these photospheric quantities instead of the effective ones.
We investigated three different model light curves denoted by A, B and C. 
The input parameters of the model A are:
$M=0.65$~M$_{\sun}$ (stellar mass), $L=$40~L$_{\sun}$ (luminosity), 
$T_{\mathrm{eff}}=6477$~K (the effective temperature of the input static model),
Z=0.0001 (metal content). For model B these parameters are: $M=0.77$~M$_{\sun}$, $L=$50~L$_{\sun}$,
$T_{\mathrm{eff}}=6300$~K, Z=0.004; and model C:  $M=0.71$~M$_{\sun}$, $L=$40~L$_{\sun}$,
$T_{\mathrm{eff}}=6300$~K, Z=0.004.

For the direct comparison we have to prepare synthetic CoRoT light curves
from the model outputs: $T_\mathrm{eff}(t)$, $\log g(t)$.
Using the basic formula of synthetic photometry \citep{Bessell05} we
determine the detected stellar fluxes in photon numbers $N_{\mathrm p}$ as
\begin{equation}\label{Np}
 N_{\mathrm p}={\frac{1}{hc}}\int{F(\lambda) \lambda R(\lambda) S(\lambda) d\lambda},
\end{equation}
where $F(\lambda)$ is the spectrum of a star in energy units, $R(\lambda)$ is
the response function of the system, $S(\lambda)$ is the filter function,
$\lambda$ is the wavelength, $h$ is the Planck constant and $c$ is the speed of light. 
The flux was computed by using the CoRoT equipment's spectral
response function $R(\lambda)$ given by \citet{Auvergne09} and without filter function
($S(\lambda)=1$).
Let us denote the result of this computation by $N^{\mathrm C}_{\mathrm p}$,
and the artificial CoRoT magnitude is 
$m_{\mathrm{CoRoT}}=-2.5 \log (N^{\mathrm C}_{\mathrm p})+c$,
where $c$ is an arbitrary constant.
We used the enhanced Kurucz models \citep{Castelli97} provided by the Spanish Virtual
Observatory\footnote{\url{http://svo.cab.inta-csic.es/theory/db2vo4/index.php?model=Kurucz}}
as theoretical flux distribution functions $F(\lambda)$. A set of model spectra
(360 synthetic spectra) were chosen which cover the physical parameters
($T_{\mathrm{eff}}$, $\log g$) of different pulsation phases and the  metallicity
range of the RR Lyrae stars, namely, $T_{\mathrm{eff}}$ is
between 5750 and 8000~K, $\log g$ 2.5 and 5.0, while [Fe/H] is
between 0 and $-$2.5~dex. Although these static model atmospheres
are not optimal for all RR Lyrae pulsation phases (see eg. \citealt{Barcza10,Barcza14})
without available, adequate dynamical model atmospheres we can use them as a first approximation.
By using $T_{\mathrm{eff}}(t)$ and $\log g(t)$ functions from the pulsation model
we assigned an interpolated $m_{\mathrm{CoRoT}}$ value to each pulsation phase. 
Generating the artificial CoRoT light curves by this way corresponds to a phase
dependent bolometric correction which was done by \citet{KovacsKanbur98}
for {\it V} band.

The Fourier solutions of the synthetic light curves are presented
in Fig.~\ref{amp_phase_modell} in the same form as the observed ones in Fig,~\ref{amp}.
The amplitude distribution of model A in top panel show similar decline to the observed 
curves, however, the higher metallicity models B and C
show double bumps. We have not seen such behaviour for any of the real stars.
It is not clear, which one of the two dips correspond to the observed single dip. 
The phase difference distributions in bottom panel of Fig~\ref{amp_phase_modell}
are phenomenological similar to the observed data (left panel in Fig.~\ref{amp}),
but the numerical $\varphi_{n1}$ values of model A are unusually high. 
We conclude that recent 1D hydrodynamic models can not reproduce the 
fine structure of the observed light curves of the RR Lyrae stars
even in the simplest non-modulated fundamental mode pulsating case.
Maybe the enhanced versions of the first successful multidimensional computations 
(2D/3D \citealt{Mundprecht13, Deupree15}) will provide us better synthetic light curves. 
For a perfect agreement between observation and theory we probably have to wait 
for sophisticated dynamical atmosphere models as well.  
We call the readers' attention to the diagrams in Fig.~\ref{amp}, which 
presents  a potentially sensitive tool  for quantitative study of the fine structures 
of RR Lyrae light curves.

\section{Estimation for physical parameters}\label{phys_par}

The standard photometric {\it V} light curves of 
RR Lyrae stars allow us to estimate their physical parameters
such as metallicity, effective temperature, $\log g$, mass, etc.
by using empirical formulae (e.g. \citealt{JK96, Jurcsik98, KW01}).
However, neither chromatic nor
monochromatic CoRoT light curves are calibrated to  standard photometric systems.
In principle, this calibration is possible, but no one has published such attempts yet.
The basic idea of such a transformation along with some of its inherent  problems is 
discussed in \citet{Weingrill15}. Similar transformation problems were 
successfully solved by \citet{Nemec11} who had to deal with 
light curves in {\it Kepler} $K_{\mathrm p}$ band.
They compared the {\it Kepler} light curves with
the existing Johnson {\it V} light curves of three non-modulated RRab stars,
then they determined numerical shifts for all useful Fourier parameters.
Using these transformed parameters and 
formulae for the {\it V} filter they received noticeably good estimates
for the physical parameters which were then verified by high-resolution
spectroscopy \citep{Nemec13}. 
A different approach was developed by the OGLE team. Since the OGLE-III RR Lyrae 
data \citep{OGLE09} are predominantly observed in Cousins $I$ colour, 
empirical formulae for {\it I} filter data were needed for physical parameter estimation. 
The OGLE team independently calibrated the empirical relations \citep{Smolec05, Pietrukowicz12} 
instead of transforming the existing formulae relevant for the {\it V} passband.
 
\subsection{Colour transformation of the CoRoT data} 

A complete calibration of CoRoT coloured and monochromatic data
is out of the scope of this paper. Instead of this we use two alternate
 transformations achieving the necessary data. 
Our primary goal is to check the physical parameters 
of our sample and to select special stars 
-- if there are any -- for further investigation. 
For this purpose a rough physical parameter estimation is enough which
can be available through these approximate transformations.

(1) Due to the direct analogy of CoRoT monochromatic and
{\it Kepler} unfiltered $K_{\mathrm p}$ data we applied the method used by
\citet{Nemec11}. Unfortunately, standard {\it V} observations of
CoRoT RR Lyrae stars are even more sparse than Kepler stars' measurements.
We found one (almost) complete phase curve for CoRoT~101370131 
in \citet{Szabo14}.  Fourier parameters were obtained
from SuperWASP data by \citet{Skarka15} for CM~Ori.
From these two data sets we derived the two most important Fourier parameters:
\begin{eqnarray}
A_1^{V}=A_1^{\mathrm{CoRoT}} + (0.10 \pm 0.04), \nonumber \\
\varphi_{31}^{V}=\varphi_{31}^{\mathrm{CoRoT}} - (0.05 \pm 0.04), \nonumber 
\end{eqnarray}
where $A_1$ denotes the Fourier amplitude of the main pulsation frequency,
$\varphi_{31}$ epoch independent phase difference is defined in
the usual way: $\varphi_{31}= \varphi_{3}-3\varphi_{1}$ \citep{Simon81},
 where $\varphi_{1}$ 
and $\varphi_{3}$ are the Fourier phases of the main frequency and the second harmonic.
The upper indices denote the used filter system.

\begin{figure}
\includegraphics[scale=0.3]{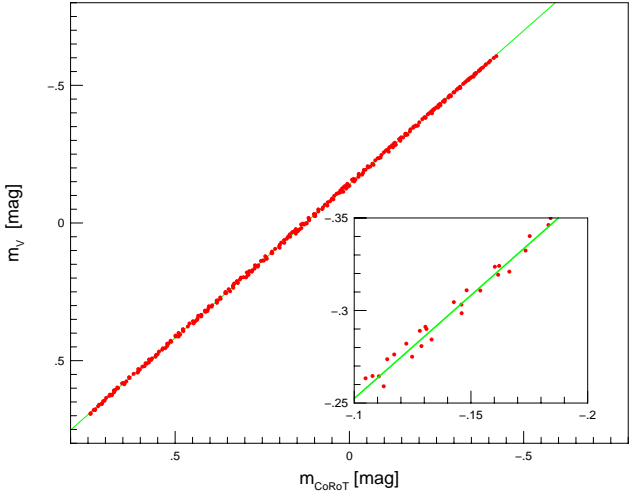}
\caption{
Correlation between the synthetic monochromatic CoRoT ($m_{\mathrm{CoRoT}}$) and 
synthetic Johnson {\it V} ($m_V$) magnitudes. The insert shows a zoom from the plot
demonstrating the weak dependence of the relation on secondary parameters (metallicity, $\log g$).}
\label{Corot-V}
\end{figure}
(2) Because of the evident weaknesses of the previous transformation 
 we alternatively applied the synthetic photometry approach 
(eg. \citealt{Straizys96, Bessell05}).
This method allow us to transform the CoRoT monochromatic RR Lyrae observations
to the standard Johnson {\it V} light curves to which we can apply the empirical
relations directly. Following the basic idea of synthetic photometry \citep{Bessell05} we
determine the detected stellar fluxes in photon numbers $N_{\mathrm p}$ as
it was shown in Eq.~\ref{Np}.
In our case, the flux was computed twice, once by using the CoRoT equipment's spectral
response function $R(\lambda)$ given by \citet{Auvergne09} and second without the filter function.
Let us denote the result of this computation by $N^{\mathrm C}_{\mathrm p}$. 
Second,  by using the same $R(\lambda)$ 
function combined with the Bessell {\it V} filter function $S(\lambda)$ \citep{Bessell90}
we get $N^{V}_{\mathrm p}$. We utilized the same set of model spectra 
(360 synthetic spectra) was used for 
the computation of synthetic CoRoT light curves in Sec.~\ref{Sec:amp_phi}.

We correlated the synthetic CoRoT magnitudes 
$m_{\mathrm{CoRoT}}=-2.5 \log_{10} N^{\mathrm C}_{\mathrm{p}} + c_1$ to the synthetic {\it V} magnitudes:
$m_{V}=-2.5 \log_{10} N^{V}_{\mathrm{p}} + c_2$. 
The average of all our  light curves transformed to the magnitude scale are set to zero, that defines a 
kind of `relative' magnitude. By using these light curves 
we avoid the need to determine the absolute zero point of the synthetic photometry which
is generally a difficult problem. 
Here we can set the constants $c_1$ and $c_2$ arbitrarily. We have chosen these so 
that the resulted magnitude ranges are similar to the observed amplitudes of RRab stars
and the zero point of the fitted line is zero.
The correlation between the synthetic CoRoT and the {\it V} magnitudes is shown in Fig.~\ref{Corot-V}.
The best-fitted linear relation is
\begin{equation}\label{fit}
m_{V} = 1.1157 (\pm 0.0008) m_{\mathrm{CoRoT}}. 
\end{equation}
The insert in Fig.~\ref{Corot-V} demonstrates the dependence of the relation on
physical parameters such as metallicity, or $\log g$. This virtual `scatter' is
an intrinsic physical property of the stars and not a scatter due to the uncertainty of our method. 
Using the ratio in Eq.~\ref{fit} we derived the transformed {\it V} light curves, from which
 we deduced the Fourier parameters (amplitudes and phase differences) necessary to use the
empirical formalae.
\begin{table*}
\begin{centering}
\caption[]{Estimated physical parameters of the CoRoT RRab stars.
The columns contain the CoRoT ID, the estimated physical parameters obtained form
empirical formulae: the metallicity [Fe/H], the absolute visual brightness $M_V$, the
reddening-free colour index $(B-V)_0$, the surface gravitational acceleration $\log g$, the effective
temperature $T_{\mathrm{eff}}$, and the mass $M$. Last column contains remarks on the Blazhko nature,
or (for star \#793) the date of the observing runs. The upper indices (1) and (2) denote the
two used colour transformation methods (see the text for the details). The
comma separated values show the results obtained
from the transformed data of method (1) and (2), respectively.
}\label{tab_par}
\begin{tabular}{@{}ccccccccc@{}}
\hline
\noalign{\smallskip}
Corot~ID &  [Fe/H]$^{(1),(2)}$ & $M_V^{(1),(2)}$ & $(B-V)_0^{(1),(2)}$ &  $\log g$ 
& $T_{\mathrm{eff}}^{(1),(2)}$  & $\log L^{(1),(2)}$  & $M^{(1),(2)}$ & rem. \\
         &         & [mag] & [mag]     &           &       [K]   &       & [M$_{\sun}$] & \\
\noalign{\smallskip}
\hline
\noalign{\smallskip}
100689{\bf 962} & $-$0.75,$-$0.65 & 0.867,0.900 & 0.314,0.324 & 3.023 & 6730,6694 & 1.519,1.524 & 0.71,0.75 &  Bl, $D_{\mathrm m} > 3$ \\
101128{\bf 793} & $-$0.75,$-$0.64 & 0.737,0.767 & 0.326,0.335 & 2.872 & 6632,6600 & 1.546,1.534 & 0.52,0.52 &  Bl, $D_{\mathrm m} > 3$ \\
101370{\bf 131} & $-$1.32,$-$1.22 & 0.539,0.568 & 0.347,0.355 & 2.728 & 6436,6407 & 1.674,1.663 & 0.57,0.57 &  $D_{\mathrm m} > 3$   \\
102326{\bf 020} & $-$0.66,$-$0.56 & 0.501,0.538 & 0.395,0.407 & 2.612 & 6248,6207 & 1.687,1.667 & 0.52,0.51 &  Bl, $D_{\mathrm m} > 3$  \\
103800{\bf 818} & $-$0.97,$-$0.86 & 0.675,0.699 & 0.305,0.312 & 2.879 & 6714,6694 & 1.578,1.564 & 0.54,0.53 &      \\
103922{\bf 434} & $-$1.22,$-$1.11 & 0.612,0.641 & 0.330,0.338 & 2.799 & 6547,6520 & 1.627,1.612 & 0.56,0.55 &   Bl \\
104315{\bf 804} & $-$0.55,$-$0.45 & 0.557,0.594 & 0.387,0.399 & 2.646 & 6304,6263 & 1.661,1.641 & 0.51,0.50 &   $D_{\mathrm m} > 3$   \\
104948{\bf 132} & $-$0.97,$-$0.87 & 0.605,0.635 & 0.344,0.353 & 2.757 & 6492,6461 & 1.635,1.619 & 0.54,0.53 &   Bl    \\
105288{\bf 363} & $-$1.13,$-$1.02 & 0.643,0.677 & 0.354,0.364 & 2.775 & 6435,6398 & 1.613,1.595 & 0.55,0.54 &   Bl \\
205924{\bf 190} & $-$1.28,$-$1.18 & 0.514,0.552 & 0.392,0.403 & 2.648 & 6210,6167 & 1.682,1.667 & 0.56,0.56 &    \\
605307{\bf 902} & $-$1.03,$-$0.92 & 0.605,0.640 & 0.367,0.377 & 2.724 & 6372,6333 & 1.636,1.623 & 0.54,0.54 &    \\
617282{\bf 043} & $-$1.39,$-$1.28 & 0.484,0.512 & 0.348,0.355 & 2.697 & 6418,6393 & 1.700,1.685 & 0.57,0.56 &    \\
651349{\bf 561} & $-$0.26,$-$0.15 & 0.702,0.740 & 0.375,0.387 & 2.735 & 6415,6371 & 1.588,1.567 & 0.49,0.48 &   Bl, $D_{\mathrm m} > 3$    \\
655183{\bf 353} & $-$1.26,$-$1.16 & 0.500,0.534 & 0.371,0.381 & 2.667 & 6312,6278 & 1.691,1.663 & 0.56,0.53 &  \\
657944{\bf 259} & $-$0.54,$-$0.43 & 0.673,0.706 & 0.353,0.363 & 2.765 & 6497,6461 & 1.601,1.582 & 0.51,0.50 &   \\
659723{\bf 739} & $-$0.89,$-$0.78 & 0.642,0.670 & 0.330,0.338 & 2.803 & 6581,6553 & 1.608,1.598 & 0.53,0.53 & 2011, $D_{\mathrm m} > 3$\\
659723{\bf 739} & $-$0.94,$-$0.84 & 0.656,0.687 & 0.337,0.346 & 2.803 & 6540,6508 & 1.604,1.588 & 0.54,0.53 & 2012, $D_{\mathrm m} > 3$\\
\noalign{\smallskip}
\hline
\end{tabular}
\end{centering}
\end{table*}

\subsection{Physical parameters}

Since no physical parameters have been derived for any of the CoRoT RR Lyrae stars, yet, 
we estimated the basic physical parameters of the complete CoRoT RRab sample. 
Only three blended stars (CoRoT 101315488, 100881648, and 101503544) were omitted, 
because of their small amplitude distorted light curves (see \citealt{Szabo14})
resulted in unusable Fourier parameters.
Although the empirical formulae were originally calibrated on the non-Blazhko stars,
since then some works showed that the formulae can be applied for well-sampled
Blazhko stars as well \citep{Kovacs05, Smolec05, Jurcsik09b, Jurcsik12, Nemec13, Skarka15}.

The calculations were done for the above (1) and (2) transformed data sets, independently. 
The metallicity was determined by using the \citet{JK96} formula,
which is defined on the \citet{CG97} scale. These metallicity values
can be transformed to the \citet{ZW84} scale as it was showed by \citet{Sandage04}.
There are signs that the formula of \citet{JK96} gives
systematically higher metallicity in the case of very low metallic content
 (see. e.g. \citealt{JK96, Nemec04, Nemec13}), 
although for example \citet{Skarka15} found no strong evidence suggesting this.
Since our sample does not contain lower metallicity stars than  [Fe/H]~$\lesssim -1.3$~dex,
we prefer the Carretta-Gratton metallicity scale as the best-fit one to the high resolution spectroscopy.
The extinction free colour index $(B-V)_0$ and $\log g$
were calculated from the formulae of \citet{Jurcsik98}, while for 
the absolute magnitude determination we followed \citet{Nemec13}, \citet{Lee14}, and \citet{Skarka15}
who applied a 0.2 zero point shift to the formula of \citet{Jurcsik98}.
The effective temperature was estimated by the formula of \citet{KW01}.

The empirical formulae for the two fundamental parameters: the stellar mass and luminosity
are yielded systematically different results depending on whether they come from the pulsation 
 \citep{Jurcsik98} or the stellar evolution \citep{Sandage06, Bono07} models. These
discrepancies were discussed in the case of {\it Kepler} RRab sample by \citet{Nemec11} in 
details. More recently, however, \citet{Marconi15} constructed consistent pulsation models
where the stellar masses and luminosities were fixed according to the horizontal-branch
evolutionary models. We took into account these new results when we computed
the masses and luminosities from the \citet{Jurcsik98} formulae than we fixed either of
these parameters and determine the other one by the van Albada-Baker type relation 
given by \citet{Marconi15} (eq.~1 in the paper).
 The typical result of this cross check is that the
pulsation mass defined by \citet{Jurcsik98} formula needs higher luminosity for
consistency than that we get from the \citet{Jurcsik98} luminosity formula. Or vice verse: \citet{Marconi15}
relation yields lower mass if we use \citet{Jurcsik98} luminosities. But in the latter case the calculated
small masses are out of the allowed parameter range where the stable pulsation persists. Therefore, we kept
the pulsation masses and determined the luminosities by the \citet{Marconi15} formulae. We mention here
that for two stars (\#793 and \#363) the alternate calculations provided consistent values for
both masses and luminosities, and for \#962 the pulsation mass (0.52~M$_{\sun}$) resulted in such 
low luminosity ($\log L$=1.416~dex) which is below the horizontal branch for the given metallicity. We accept  the evolutionary mass and pulsation luminosity for \#962. 

The results are summarized in Table~\ref{tab_par}. 
In spite of the rather ad hoc character of the two colour transformations 
the estimated parameters are surprisingly consistent. The differences between 
the comma separated columns (obtained from the two transformations) indicate the 
accuracy of the estimated parameters.
We checked the consistence of the light curve shapes with the calibration sample of
the empirical formulae using the $D_{\mathrm m}$ parameter \citep{JK96}.
Several Blazhko stars yield $D_{\mathrm m}$<3 values which supports that their
estimated physical parameters should be as good as the non-Blazhko stars' ones.
There is a rather surprising result of this test: for three non-Blazhko
stars (\#131, \#804, and \#739) $D_{\mathrm m}$>3 which suggests some light curve distortion. 
We compared the observed individual Fourier parameters of these stars with
the calculated ones by the interrelation of \citet{JK96}. We found that
the observed amplitudes (e.g. $A_1$) are always smaller that the predicted ones.
This implies some flux loss for these stars. This would not be unprecedented.
Similar problems have been discussed for several {\it Kepler} RR Lyrae stars \citep{Benko14}. 
The explanation of the possible flux lost is simple: the star fluxes are collected within
pre-defined pixel masks for both CoRoT and {\it Kepler}. In the preparatory phase
of the missions photometric sky surveys had been conducted and the brightness values of the input catalogues
(Exo-DAT, KIC) define the size of the masks. Since these missions are optimized to
detect small brightness variations (transiting exoplanets, solar type oscillations)
this strategy is generally appropriate. However, the RR Lyrae stars have large amplitudes
with highly non-sinusoidal light curve shapes, viz. they spend most of their time in faint phases.
If a mask is fitted to this faint state, it is too small around maximal brightness resulting 
in flux loss. This hypothesis is strengthened by the fact that
these problematic stars (\#131, \#804, and \#739) are the faintest (15.6, 15.94, and 15.186~mag)
non-Blazhko RR Lyrae stars in the Exo-DAT catalogue.
Additionally, the star \#739 was observed using different masks for the two
 runs (LRc07 and LRc10). The light curves of the two runs are different, e.g. they
show different amplitudes and other Fourier parameters are also slightly different.
If both masks contained the total flux no such differences would be expected.

\begin{figure}
\includegraphics[scale=0.3]{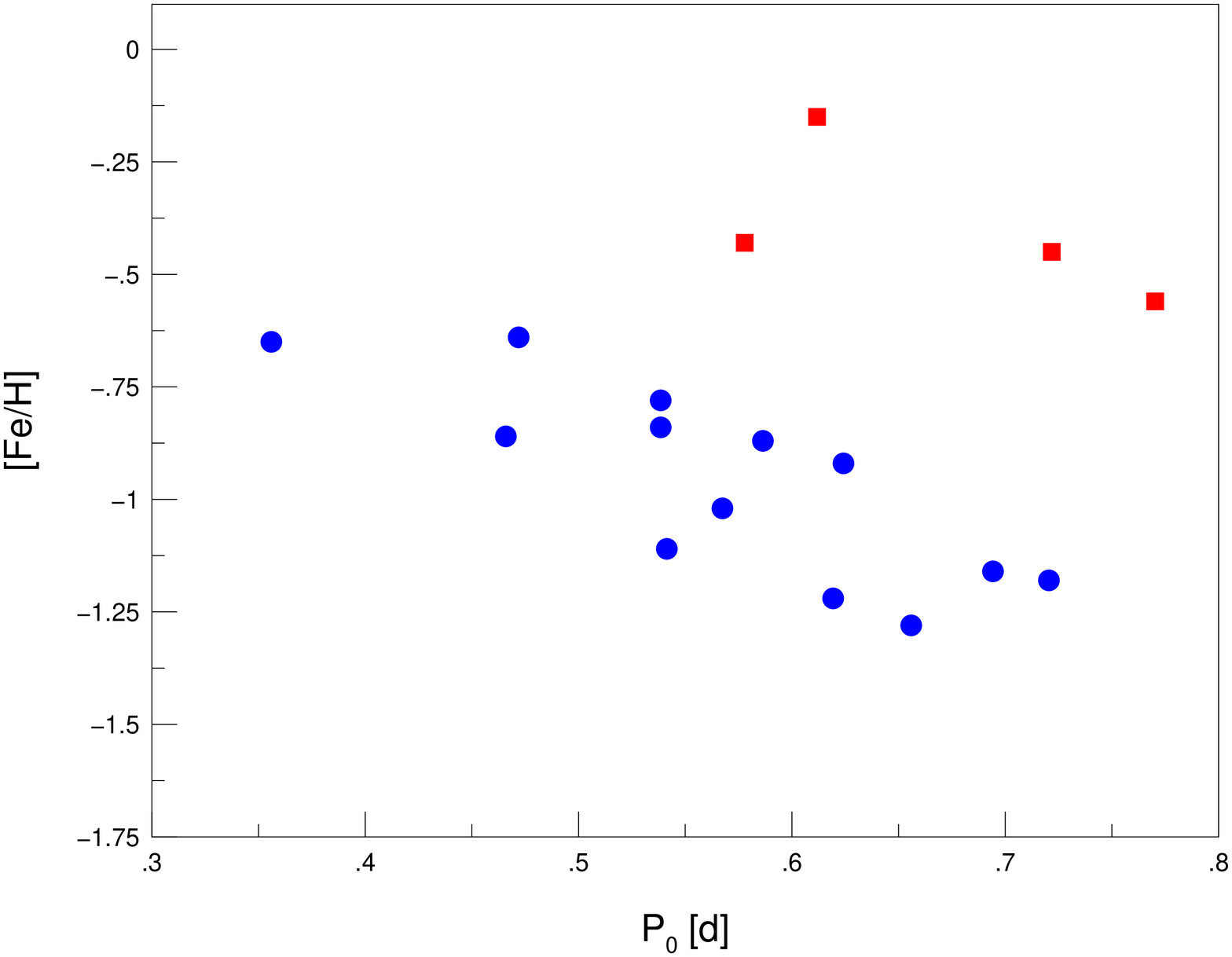}
\caption{
Correlation between the pulsation period $P_0$ and the metallicity
[Fe/H]$^{(2)}$. The blue dots define a moderate metallicity stellar group
with the average metalliciy of <[Fe/H]>=-0.96~dex, while red squares 
show a higher metallic content group with <[Fe/H]>=-0.39~dex.
The red square between the two groups shows the star \#259 with <[Fe/H]>=-0.43~dex.
}
\label{FeHP}
\end{figure}
The original goal of this parameter estimation was to find any extreme
stars in our sample. The physical parameters of all stars are, however, in the canonical
ranges of the equilibrium values of RRab stars. The metallicity is 
between $-0.15 > \mathrm{[Fe/H]} > -1.39$, the absolute visual brightness 
is $0.9 > M_{V} > 0.484$~mag, the reddening free color index is
$0.305 > (B-V)_0 >0.407$, the effective 
temperature is $6210 > T_{\mathrm{eff}} >6730$~K,
the surface gravitational acceleration is $3.023 > \log g > 2.612$,
the luminosity is $1.700 > \log L >1.519$, and the mass is $0.75 > M > 0.48$~M$_{\sun}$.  
By reviewing some classical diagnostic diagrams of RR Lyrae stars
such as period vs. amplitude, $(B-V)_0$ vs. $M_V$, etc. we found
that the CoRoT sample consists of two distinct groups.
The separation of the two groups is the best seen in 
period vs. metallicity diagram (see Fig.~\ref{FeHP}). The blue symbols
of the figure define a moderately metal poor group with an average
metallicity of <[Fe/H]>=$-$0.96~dex, while the red filled squares denote 
the most metal rich stars of the sample 
(\#020, \#804, \#561, <[Fe/H]>=$-0.39$~dex) and
the star \#259 which is located  at an intermediate position between of these two
groups. The metal rich group has significantly higher average 
period <$P_0$>=0.7013~d than the more populous metal poorer group 
<$P_0$>=0.5677~d. These stars are at the red edge 
when we plot the sample in the theoretical HR diagram
($\log T_{\mathrm{eff}}$, $\log L$). This is not a surprise,
since the instability strip moves red-ward with the increasing metallicity.

\section{Summary}\label{sec:concl}

We presented here the results of a rigorous RR Lyrae search in the 
CoRoT archive. We found nine RRab stars in the data base that have not been studied and 
seven of them are new discoveries. 

Three stars show the Blazhko effect. The cycle-to-cycle variation of these Blazhko effects are evident 
either for amplitude or frequency modulations or both which rises the possibility 
of their multiperiodic (or irregular/chaotic) nature.
The Fourier spectrum of star \#132 contains small amplitude `additional'
frequencies which can be identified as the 
consequence of the period doubling effect and the excitation of the 
second radial overtone mode. The later mode identification is strengthened by 
pulsation modeling, since the observed period ratio is unusually 
high ($P_2/P_0=0.601$), but linear pulsation models at least do not exclude this period ratio.
The other two Blazhko stars show no significant additional frequencies.

The light curve stability and period changes were studied for non-Blazhko
stars. We detected a significant cycle-to-cycle fluctuation of the 
pulsation period (cycle length) of CM~Ori. This is the first case that 
random jitter has been detected for an RR Lyrae star. The fluctuation rate 
is tiny: at longest 1-2 seconds per cycles. 

Long term period changes were also investigated for the two known RR Lyrae
stars CM~Ori and V2042~Oph. Their O$-$C diagrams cover more than 80 and almost 70 years,
respectively. Both stars show a  slight period increase with rates which agree well
with the prediction of the stellar evolution theory for normal red-ward evolution.

The Fourier amplitude and phase difference distribution with
respect to the harmonic order have been studied for the first time on
a larger set of non-Blazhko stars. We found a common amplitude decline feature.
The phase differences show a split distribution. 
The observed distributions were compared theoretical works.
Our conclusion is that the recent model calculations reproduce the
main features but the fine structures can not be described properly.
These amplitude and phase distribution 
diagrams could be potential diagnostic tools for constraining  pulsation models,
because of their high sensitivity of the fine structure of the light curves.

We successfully transformed the CoRoT unfiltered light curves, and
their Fourier parameters to the Johnson {\it V} curves, and parameters.
As a by-product of the use of the interrelations we found possible 
flux loss for several stars. Such an effect was shown for high amplitude
variables of the {\it Kepler} telescope, but has not documented
yet in the case of CoRoT satellite.
         
By using empirical formulae we
estimated the basic physical parameters of the complete CoRoT RRab sample.
The estimated physical parameters define two subgroups of the sample.
(1) A shorter period and lower metallicity and (2) a longer period higher
metallicity groups. The physical parameters of both groups, however,
are within the canonical ranges of RR Lyrae stars.

\section*{Acknowledgements}

This work was supported by the ESA PECS Grant No
4000103541/11/NL/KML and the Hungarian National Research Development and 
Innovation Office -- NKFIH K-115709. \'AS was supported by the J\'anos Bolyai 
Research Scholarship of the Hungarian Academy of Sciences. 
The research leading to these results has received funding from the
European Community's Seventh Framework Programme (FP7/2007-2013) under grant agreement no. 312844.






\bsp	
\label{lastpage}
\end{document}